\documentclass[preprint]{emulateapj}
\slugcomment{Accepted for Publication in the Astrophysical Journal Letters}
\shortauthors{Zaritsky et al.}
\shorttitle{Chemical Enrichment of the Intracluster Medium}

\begin{document}
\title{
Intracluster Stars and
 the Chemical Enrichment of the Intracluster Medium
}
  
\author{Dennis Zaritsky\altaffilmark{1}, Anthony H. Gonzalez\altaffilmark{2,}\altaffilmark{3},
and Ann I. Zabludoff\altaffilmark{1}}
\altaffiltext{1}{Steward Observatory, University of Arizona, 933 North Cherry Avenue, Tucson, AZ 85721}
\altaffiltext{2}{NSF Astronomy and Astrophysics Postdoctoral Fellow}
\altaffiltext{3}{Department of Astronomy, University of Florida, Gainesville, FL 32611}
\email{dzaritsky@as.arizona.edu, anthony@astro.ufl.edu,azabludoff@as.arizona.edu}

\begin{abstract}         
We explore the contribution of intracluster stars (ICS) to the
chemical enrichment history of the intracluster medium (ICM). 
In contrast to scenarios in which all the metals originate in cluster
galaxies and are then transported into the ICM,  
intracluster stars 
enrich the ICM {\sl in situ}, thereby contributing
100\% of their supernovae ejecta directly into the ICM. Modeling the ICS as 
an ancient, single burst stellar population with
a normal initial mass function, we generate ICM iron abundances
in the range of  the observed values of several tenths solar. 
Large observational and theoretical uncertainties preclude us from concluding that
the intracluster stars are the primary contributor of metals to the ICM in general. 
However, for the two clusters in our sample, and one from the literature,
for which all of the required observational constraints exist, we
are able to reproduce between half and all of the ICM iron with this simple model.
Due to the ubiquity of intracluster stars in
clusters and their direct connection with the ICM, we conclude that
all models of the chemical enrichment history and energy budget of the ICM
should account for the impact of the ICS.
\end{abstract}

\keywords{cosmology:observation --- galaxies:clusters:general --- supernovae}

\section{Introduction}
\label{sec:intro}

The chemical enrichment history of the intracluster medium (ICM) 
is arguably the least understood aspect of galaxy cluster evolution.
Particularly puzzling are the relatively high Fe abundance \citep[$\sim 0.3$ Fe$_\odot$;]
[]{edge} and moderately large $[\alpha/$Fe$]$ \citep{renzini93, mush96, tamura}. 
The former
has been interpreted as evidence for large gas outflows from 
galaxies \citep{renzini93, renzini97}, while the latter, in combination with constraints
from cluster galaxies, is interpreted as evidence for
a non-standard initial mass function \citep[][although see \cite{pipino} for 
a contrasting result]{mush96, portinari}. Both of these inferences
have a wide ranging impact on a variety of astronomical topics. Here we present
evidence that the first argument should be significantly revised, and 
that the second argument should be revisited.

Our proposal is straightforward. The intracluster stars (ICS) that we have found without
exception in a sample of 24 clusters \citep[][hereafter Paper I]{paper1} must chemically pollute the
ICM via supernovae. Such intergalactic SNe have already been observed \citep{galyam}. We evaluate the degree to which the ICS plays a role with the simplest modeling 
we can envision.  As the
vast literature on this topic demonstrates, there are many variables and unknowns in this 
type of model. We will adopt {\sl plausible}, often 
typical, values of the relevant variables and demonstrate that the
intracluster stars make a significant, and perhaps dominant, contribution to the enrichment history
of the ICM. This study is not intended to be a complete or comprehensive treatment
of the ICS enrichment of the ICM, but it  illustrates
the likely importance of the ICS in the chemical abundance budget of galaxy clusters.

\section{Observational Constraints on the Intracluster Components}
\label{sec:data}

In Paper I we present observations of 
24 clusters that span a range of velocity dispersions
and Bautz-Morgan types \citep{bautzmorgan1970}. 
The sample consists of nearby clusters ($0.03<z<0.13$)
that contain a dominant brightest cluster galaxy (BCG) with a 
major axis position angle that lies within 45 degrees of the east-west
axis (our drift scan direction). We present details of these unique data and our reduction
procedure in Paper I.

From modeling the two dimensional optical surface brightness distribution, we
conclude in Paper I that the BCG light can be divided into two components. The 
inner component is similar to normal giant ellipticals in extent, luminosity, and scaling relations.
The outer component is closely related
in scale and ellipticity to the distribution of galaxies in the cluster.
We associate the inner component with the 
BCG and the outer component with the ICS.
In Paper I, we determine the luminosity of the outer
component, which we will use here as a measure of the intracluster stellar content.
Standard measurements of BCG light underestimate the contribution of this
low surface brightness outer component
(for example, assuming a single $r^{1/4}$ surface brightness profile
to evaluate the total magnitude leads to a 50\% underestimate of the total luminosity of
the BCG plus ICS; Paper I).

For constraints on the properties of the ICM,
we have searched the literature for
X-ray gas masses \citep{reiprich} and chemical abundances \citep{white}. 
There are only
two clusters  in our sample for which such data exist (Table \ref{tab:obs}).
We have adopted the gas mass inside of $r_{500}$, the radius at which the
mean mass density exceeds 500 times the cosmological value, as the measure
of the gas mass. We measure the ICS luminosity directly only within 300 kpc, but our
model fit provides an estimate of the total luminosity. The
mismatched apertures limit the precision with which comparisons can
be made, although even if the apertures are matched, the comparison
is complicated by flows of material across the aperture during the cluster's evolution. 
The gas masses have been rescaled to our adopted cosmology
($H_0 = 70$ km/sec/Mpc, 
$\Omega_m = 0.3$ and $\Omega_\Lambda = 0.7$), and the observed
abundances have been rescaled due to the difference between the older
value of the photospheric solar abundance by mass \citep[0.0026,][]{anders89}, and 
the more recent one that we adopt here \citep[0.0017,][]{grevesse}. This
conversion increases the canonical Fe cluster abundance from 0.3 to 0.46 Fe$_\odot$.

\section{Models and Results}
\label{sec:models}

This type of study usually relies on analytic models of the enrichment
history \citep{renzini93, bm, maoz, portinari}, but we choose 
to model the chemical evolution 
with the publicly available modeling package, PEGASE.2 \citep{pegase},
which provides estimates of the Type I and II supernovae rates as a 
function of age for a user-selected star formation history
and initial mass function. 
We use these models because they also provide
luminosities and colors of the resulting stellar population, which we need
to scale our predictions for the observed ICS population. 

To mitigate ambiguity
in our results, we adopt
a very simple star formation history (a single burst of star formation
that peaks at $z > 10$ and is finished by $z = 4$), an initial chemical
abundance by mass fraction of 0.004 for the ISM (although adopting values
of 0.0001 and 0.08 does not significantly affect the results),
a standard IMF \citep{scalo}, and all the default PEGASE parameters (such
as a binary fraction of 0.5). Changing the star formation history so that
the peak star formation rate is at $z =$ 3.5 and the star formation does not
cease entirely until $z \sim 2$ decreases the predicted abundances by 
$\sim 10\%$. Varying the adopted initial chemical abundance only affects aspects of
the stellar evolutionary modeling, such as the predicted stellar colors. We
always assume that the ICM abundance is zero until we add the contribution from 
the ICS.

\begin{figure}
\figurenum{1}
\plotone{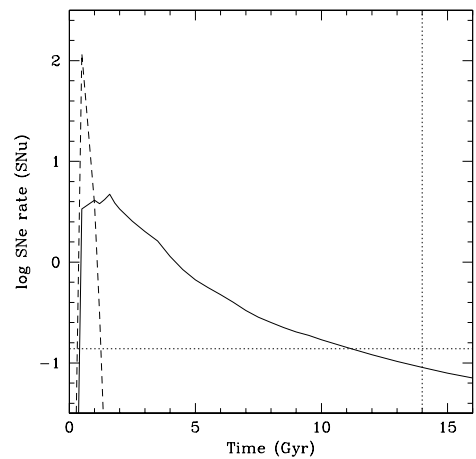}
\figcaption{SNe rates for Type I (solid line) and II (dashed
line) over time from the PEGASE models. The units are SNu (SNe per 10$^{10}$ L$_{B,\odot}$ per 
century). The horizontal dotted line represents the SNe rate observed
locally \citep{cappellaro} and the vertical dotted line represents the
current time. For the adopted parameters, the PEGASE models
produce a slightly
conservative (lower) value of the SNe rate at the current time than is observed.
}
\label{fig:sne}
\end{figure}

By integrating the rate of SNe predicted from the PEGASE models (Figure 
\ref{fig:sne}) and adopting Fe yields for SNe, we obtain the total
Fe mass ejected over time. For the yields we assume that
a Type I ejects 0.7$M_\odot$ of Fe per supernova and that a 
Type II ejects 0.05$M_\odot$ of Fe per supernova. This yield for
a Type I is the canonical value adopted by various other 
studies \citep[see][]{maoz, portinari}, while that for Type II's
is the mean value \citep{elmhamdi} over the observed range, 0.0016 to 0.26 $M_\odot$ \citep{hamuy}. Because Type I's
dominate the enrichment in our models, the results are fairly 
robust to decreasing the Type II yield from our adopted value. Increasing the
Type II yield leads to more pronounced changes. However, increasing the Type II yield 
obviously produces a larger total Fe mass and therefore is not a conservative
approach. 

This discussion sidesteps the difficulties others have encountered
in reproducing $[\alpha$/Fe$]$ or specific element ratios \citep{renzini93, mush96}.
However, any solution that has been  proposed previously, such as
non-standard initial mass functions, 
will also work with our ICS scenario, with the added flexibility that now one
does not need to fit both the ICM and galaxy abundance patterns with the same
model. Therefore, although we do not pursue the [$\alpha$/Fe] question here, we
conclude that it is no more difficult a problem than in the standard models. 
Future models should be able to take advantage of the wealth of information
becoming available on the spatial gradients of these element ratios \citep{tamura}.

\begin{figure}
\figurenum{2}
\plotone{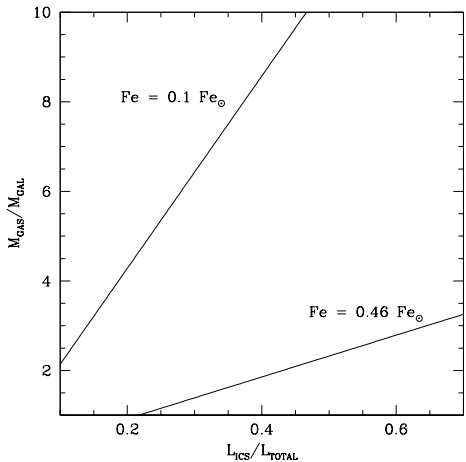}
\figcaption{Parameter choices that produce ICM metallicities of 0.1
and 0.46 Fe$_\odot$ are shown by the lines (an abundance of 0.46 Fe$_\odot$ corresponds
to the canonical $\sim 0.3$ abundance \citep{edge} for our adopted
photospheric Fe solar abundance). The vertical axis represents the mass ratio of
the ICM to stars within galaxies. The horizontal axis represents the fraction of all
cluster stars that are in the ICS. Although producing 
an ICM metallically of 46\% solar strains the observational
constraints on both axes (see text), 
producing a significant contribution, $\sim 0.1$ to $0.2$ Fe$_\odot$,
is well within the observational limits. Given the modeling and observational
uncertainties, we cannot exclude the possibility that the bulk of the ICM Fe comes
from the ICS.
}
\label{fig:ratios}
\end{figure}

We present  two different calculations that provide  
quantitative estimates of the importance of the ICS.
First, we model clusters as a general population,
adopting typical ratios of such quantities as gas mass to stellar mass.
Second, we model the only three clusters for which all of the required quantities 
are measured. In both approaches, we 
combine the PEGASE supernovae 
rates with the yields described above to calculate the Fe mass that 
1 M$_\odot$ of initial material produces after 14 Gyr (this 
1 M$_\odot$ of material results in 0.59 M$_\odot$ of stellar mass,
including remnants, at the current time). In the first type of calculation, we utilize the ratios
of gas mass to galaxy mass in stars ($M_{GAS}/M_{GAL}$)  and of ICS to total cluster
luminosity ($L_{ICS}/L_{TOTAL}$), and 
assume that that the ICS and galactic stars have the same M/L ratio,  to
calculate the ratio of ICM gas mass to ICS mass. We then use this ratio,
and the calculated Fe abundance per solar mass of ICS, 
to calculate an ICM Fe abundance.
Values for $M_{GAS}/M_{GAL}$ vary between $\sim 2$ to 10
\citep{arnaud,fukugita}, although a combination of recent optical work with
the Sloan Digital Sky Survey \citep{bell}
and a large ROSAT sample \citep{ettori} favors the upper end of that range
\citep{mushotzky04}. In the limited observational work to date, values for $L_{ICS}/L_{TOTAL}$
range from 0.1 \citep{feldmeier2004} to 0.2 \citep[][adjusted for the division into
two components of Abell 1651 from Paper I]{paper0}.  If, as models predict \citep{sl, murante},
the stars in the ICS are on average older than those in the galaxies, then our assumption of 
equal M/L for the ICS and galactic stars leads to an underestimate
of the relative amount of mass in the ICS. This is therefore a conservative assumption in 
our determination of the produced Fe. 
Instead of adopting specific values for the various ratios, we provide results
over the range of plausible values (Figure \ref{fig:ratios}). 

Figure \ref{fig:ratios} demonstrates that for the ICS to chemically
enrich the ICM to cluster-like values
requires both a fairly high $L_{ICS}/L_{TOTAL}$  and
a low $M_{GAS}/M_{GAL}$  relative to the
ranges of values explored. 
However, to reach a level where the ICS contributes significantly to the ICM
($\sim 0.1 - 0.2$ Fe$_\odot$) requires values of the two scaling ratios that lie well within the observational ranges
\citep{arnaud, paper0, feldmeier2004}. Furthermore, because 
M/L is likely to be
larger for the ICS than for the galaxies (see above),
our scaling may underestimate the contribution from the ICS by 
up to a factor of two.

The second type of  calculation provides a
more direct comparison in the few clusters for which
the relevant data are available.
For two of our clusters,
we find corresponding $M_{GAS}$ and metallicity
measurements in the literature (Table \ref{tab:obs}). 
We scale the Fe mass, as calculated 
from the PEGASE models and our adopted yields, 
using the M/L from those models and
the total luminosity of the outer component (Paper I). The ICM without the 
contribution of the ICS is assumed to have zero metallicity. The calculated
metallicity resulting from the ICS contribution is
presented in Table \ref{tab:obs} for comparison. $M_{ICS}$ represents the
stellar mass, including remnants, in the intracluster volume.

\begin{deluxetable*}{lrrrrrrrr}
\tabletypesize{\scriptsize}
\tablecaption{Cluster Parameters and Estimates of the Intracluster Fe Abundance}
\tablewidth{0pt}
\tablehead{
\colhead{Cluster}&
\colhead{$z_{BCG}$}&
\colhead{M$_{gas,500}$}&
\colhead{r$_{500}$}&
\colhead{M$_I$}&
\colhead{M$_V$}&
\colhead{M$_{ICS}$}&
\colhead{Observed Fe}&
\colhead{Predicted Fe}\\
\colhead{}&
\colhead{}&
\colhead{($10^{12} M_\odot$)}&
\colhead{(Mpc)}&
\colhead{}&
\colhead{}&
\colhead{($10^{12} M_\odot$)}&
\colhead{(solar units)}&
\colhead{(solar units)}
\\
}
\startdata
Abell 1651  & 0.0853 & 6.5 &1.73&$-$24.95 & ...& 0.9&0.44&0.22\\
Abell 1914$^a$\tablenotetext{a}{This cluster is drawn from
 \cite{feldmeier2004}. Their measurement of the ICS is somewhat different than ours,
partly because they observe clusters without a dominant brightest member.
We adopt the ICS luminosity corresponding to their mask threshold, which best 
approximates our approach. Their measurement is in the $V$ band, which further
complicates the comparison.}&0.1712&12.9&2.46&...&$-$24.94&1.8&0.37&0.22\\
Abell 3112  & 0.0759 & 4.5 &1.53&$-$25.74 & ...&1.9&0.68&0.66\\
\enddata
\label{tab:obs}
\end{deluxetable*}

It is unfortunate that only two of our 24 clusters have published values of
$M_{GAS}$ and Fe abundance. However, for those two clusters, we find that the ICS
can account for a large fraction of the metals in the ICM, and we 
reproduce the relative rankings of metallicity for the two clusters. 
There are numerous 
adjustments that can be made due to uncertainties 
in the luminosity of the ICS, the region over which $M_{GAS}$ and the ICS 
are measured, the adopted yields, the modeling of the SNe rates,  the star
formation history, initial mass function, and so forth. In Table \ref{tab:obs}
we also include results for a cluster observed by \cite{feldmeier2004}. Despite
some differences in their definition and measurement of the ICS, our model
again reproduces a sizable fraction of the ICM iron abundance in this cluster.
The purpose of this exercise is primarily to demonstrate that for 
plausible values of the various parameters, the ICS can contribute
a significant amount of iron to the ICM. A similar conclusion has been reached
using scaling relationships and infrared photometry by \cite{lm}.

A complete model of cluster enrichment must include the contribution from
all stars, whether currently in the ICS or in galaxies, and must account for
the metals both in the ICM and in galaxies. Our approach has ignored 1) the
contribution of metals to the ICM from stars currently in galaxies, because we are interested 
in determining the contribution from the newly measured ICS component, and
2) the metals currently locked in galaxies, again because we are interested
in the connection between the ICS and ICM. In effect, we have treated the
current cluster galaxies as closed box systems and set them aside.
This approach produces a conservative estimate of the metal enrichment of the ICM.

It is important to discuss one possible refinement of our chemical enrichment
model.  We make no direct assumption about the origin of the ICS, which, in simulations, 
arises from the tidal disruption of cluster galaxies early in the cluster's history
\citep{sl,murante,willman2004}.  If these simulations are correct, then the metals in the ICM are
some combination of the metals already present in those first cluster galaxies and
the subsequent evolution of the stars stripped from those galaxies.  
So far in this paper, we have discussed only the latter issue. 

We now consider the
former issue by asking what fraction of the metals in those ICS progenitor galaxies 
enters the ICM.
If the parent galaxies of the ICS were completely disrupted,
then the answer is 100\% and our model implicitly includes this possibility. If some of the
parent galaxies survive to the current time, then our calculation overestimates the
ICS metal contribution to the ICM by whatever amount of metals is locked in these galaxies.
Without knowing when the ICS formed and its detailed star formation history, we cannot
estimate the magnitude of this effect, but it may be significant. 
For example, for a star formation model that peaks sharply at $z \sim 3$,
an age comparable to the mean age of the ICS in the simulation by \cite{sl},
about 25\% of the Fe forms prior to the mean age of the ICS stars.
If the stars are stripped from these galaxies shortly thereafter, the galaxies 
might retain as much as 25\% of all the metals that we attribute to the ICS.
Nevertheless, even if our calculation overestimates the
ICS metal contribution to the ICM by as much as 25\%, the ICS still
contributes between 35\% and 75\% of the metals in the three clusters that we
model in detail (Table 1).

\section{Conclusions}
\label{sec:discussion}

We calculate that the chemical
enrichment due to intracluster stars (ICS),
a ubiquitous population in galaxy clusters (Paper I), is a significant component of the
chemical enrichment history of the intracluster medium (ICM). 
Furthermore, our model
suggests that for at least some clusters the ICS could contribute the bulk
of the metals in the ICM.
Unlike models of 
ICM enrichment in which the metals originated in the cluster members
we see today, the simplest version of our model has no
uncertainty related to the fraction of the processed material from 
the ICS that pollutes the ICM. In fact, once the relevant measurements (ICM mass, ICM metallicity,
and ICS mass) are made sufficiently
precise, our argument could be turned around to place limits on the metal
outflow from cluster galaxies. 
The ICS, and the associated SNe, should also be an important
factor in the energy budget of the ICM (see \cite{bm} for such
calculations relating to the SNe within the cluster galaxies and \cite{domainko}
for the effects of ICS SNe on large-scale cluster gas dynamics). Lastly, 
Table \ref{tab:obs} suggests that the factor of $\sim$ 2 scatter in abundances 
among clusters \citep{baum, horner} is
related to the balance between the mass of the ICM gas and that of  the ICS. 
We speculate that the trend seen in Fe abundance
versus temperature \citep{horner01, buote} may reflect an interplay between the
ICS fraction, which appears to decrease in less massive clusters
\citep{feldmeier2004}, and the gas fraction, which varies in a complex
manner as a function of temperature
\citep{sanderson}.

This paper is not
a detailed exploration of the parameter space currently accessible to models
of ICM enrichment by the ICS, but
rather a demonstration case for the importance of including the
ICS in any accounting of the chemical enrichment 
history of clusters. As such, our conclusions are based on simple models that adopt
typical values of such unknown quantities as the initial mass function and stellar
yields. Future work should aim to obtain the requisite observational
constraints to test these models in detail across a larger
sample of clusters.

\begin{acknowledgements}
We thank Richard Mushotzky for comments on a preliminary draft and Wayne
Baumgartner for providing data prior to publication. We also thank Yen-Ting Lin
and Joe Mohr for communicating their results prior to publication.
DZ acknowledges support from a David and Lucile Packard Foundation Fellowship
and NASA grant NAG5-13583.
AIZ is supported by NSF grant AST-0206083 and NASA LTSA grant NAG5-11108.
AHG is supported by an NSF Astronomy and Astrophysics Postdoctoral 
Fellowship  under award AST-0407085. 
\end{acknowledgements}

\end{document}